# Optimal Data Structures for Farthest-Point Queries in Cactus Networks[*]

Prosenjit Bose[†]    Jean-Lou De Carufel[†]    Carsten Grimm[†‡§]
Anil Maheshwari[†]    Michiel Smid[†]

November 7, 2014

### Abstract

Consider the continuum of points on the edges of a network, i.e., a connected, undirected graph with positive edge weights. We measure the distance between these points in terms of the weighted shortest path distance, called the *network distance*. Within this metric space, we study farthest points and farthest distances. We introduce optimal data structures supporting queries for the farthest distance and the farthest points on trees, cycles, uni-cyclic networks, and cactus networks.

## 1 Introduction

Consider the continuum of points on the edges of a network, i.e., a graph with positive edge weights. We measure the distance between these points in terms of the weighted shortest path distance, called the *network distance*. Within this metric space, we study farthest points and farthest distances.

Decisions where to place facilities are often complex and involve optimization of multiple criteria. Our data structures enable decision makers to quickly compare farthest distances from potential locations, which may constitute an essential factor, e.g., impacting emergency response times for a possible location of a new hospital. Furthermore, our results provide a *heat-map* of farthest distances illuminating the aspect of centrality of the network at hand and, thereby, serve as a visual aid for decision makers.

We introduce optimal data structures supporting queries for the farthest distance and the farthest points on trees, cycles, uni-cyclic networks, and cactus networks. We begin with data structures for simple networks and then use them as building blocks for more complex networks. With this modular approach we can easily extend our results as new building blocks become available.

The remainder of this section is organized as follows. In Section 1.1, we set the stage for this work by making our notions of networks, points along networks, and network distance precise. In Section 1.2, we summarize related work. In Section 1.3, we outline our main contributions and the structure of this work.

---

[*]This research has been partially funded by NSERC and by a fellowship from the German Academic Exchange Service (DAAD). A preliminary version of this work was presented at the 25th Canadian Conference on Computational Geometry [3] and was part of the Diplomarbeit (Master's thesis) of the third author [9].

[†]School of Computer Science, Carleton University
[‡]Institut für Simulation und Graphik, Fakultät für Informatik, Otto-von-Guericke-Universität Magdeburg
[§]Corresponding author carsten.grimm@ovgu.de

## 1.1 Preliminaries and Problem Definition

We call a simple, finite, undirected graph with positive edge weights a *network*. Unless stated otherwise, we consider only connected networks. Let $G = (V, E)$ be a network with $n$ vertices and $m$ edges, where $V$ is the set of vertices and $E$ is the set of edges. We write $uv$ to denote an edge with endpoints $u, v \in V$ and we write $w_{uv}$ to denote its weight. A point $p$ on edge $uv$ subdivides $uv$ into two sub-edges $up$ and $pv$ with $w_{up} = \lambda w_{uv}$ and $w_{pv} = (1 - \lambda) w_{uv}$, where $\lambda$ is the real number in $[0, 1]$ for which $p = \lambda u + (1 - \lambda) v$. We write $p \in uv$ when $p$ is on edge $uv$ and $p \in G$ when $p$ is on some edge of $G$.

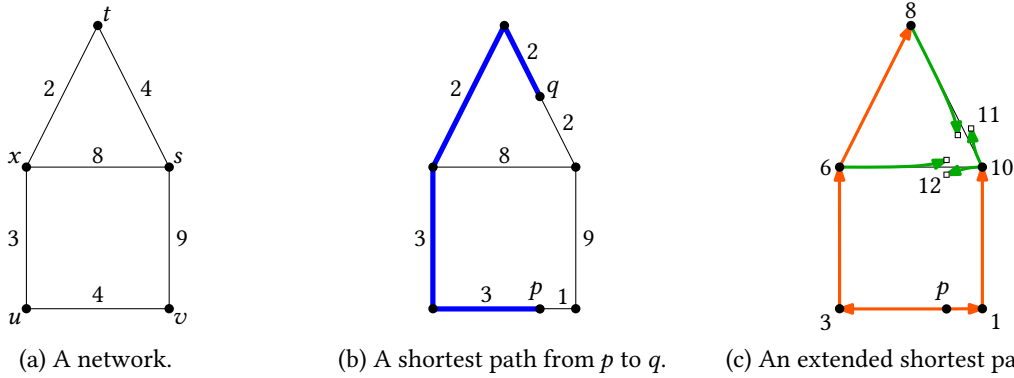

(a) A network.      (b) A shortest path from $p$ to $q$.      (c) An extended shortest path tree.

Figure 1: In the network shown in (a), the network distance from $p = \frac{1}{4}u + \frac{3}{4}v$ to $q = \frac{1}{2}s + \frac{1}{2}t$ is $d(p, q) = 10$. This distance is, for instance, achieved along the shortest path (blue) from $p$ to $q$ depicted in (b). The eccentricity of $p$ is $\text{ecc}(p) = 12$ and the farthest point from $p$ lies along the edge $xs$, as indicated by the shortest path tree from $p$ (orange) and its extension (orange ∪ green) illustrated in (c).

As illustrated in Figure 1, we measure distance between points $p, q \in G$ in terms of the weighted length of a shortest path from $p$ to $q$ in $G$, denoted by $d_G(p, q)$. We say that $p$ and $q$ have *network distance* $d_G(p, q)$. The points on $G$ and the network distance form a metric space. Within this metric space, we study farthest points and farthest distances. We call the largest network distance from some point $p$ on $G$ the *eccentricity* of $p$ and denote it by $\text{ecc}_G(p)$, i.e., $\text{ecc}_G(p) = \max_{q \in G} d_G(p, q)$. A point $p$ on $G$ is farthest from $p$ if and only if $d_G(p, \bar{p}) = \text{ecc}_G(p)$. We omit the subscript $G$ whenever the network is understood from the context.

We extend the definition of shortest path trees from graphs to networks. When we say we *cut* an edge $ab$ at a point $p \in ab$ we mean that we introduce two new vertices $x$ and $y$ and replace the edge $ab$ with two edges $ax$ and $yb$ of weight $w_{ax} = w_{ap}$ and $w_{yb} = w_{pb}$. If $p$ coincides with an endpoint of $ab$, then one of $ax$ and $yb$ has weight zero and is omitted. Let $s$ be some point on a network $G$, and let $T_s$ be the shortest path tree of $s$ in $G$. We split each non-$T_s$ edge $ab$ at the farthest point from $s$ on $ab$ and add the resulting edges $ax$ and $yb$ to $T_s$. The resulting tree is called the *extended shortest path tree* [25] of $s$ in $G$; this tree encapsulates both the eccentricity of $s$ in $G$ and the farthest points from $s$ in $G$, as illustrated in Figure 1c.

We aim to construct data structures for a fixed network $G$ supporting the following queries. Given a point $p$ on $G$, what is the eccentricity of $p$? What is the set of farthest points from $p$ in $G$? We refer to the former as an *eccentricity query* and to the latter as a *farthest-point query*. Both queries consist of the query point $p$ represented by the edge $uv$ containing $p$ and the value $\lambda \in [0, 1]$ such that $p = \lambda u + (1 - \lambda) v$.

We study trees, cycles, uni-cyclic networks, and cactus networks. A *uni-cyclic network* is a network with exactly one simple cycle. A *cactus network* is a network in which no two simple cycles share an edge or, in other words, a network where each edge is contained in at most one simple cycle.

## 1.2 Related Work

Our data structures implicitly represent (generalized) farthest-point network Voronoi diagrams, where the sites are the entire continuum of points along a network [4]. Usually, Voronoi diagrams are defined with respect to a finite set of sites representing points of interest in some metric space [1, 24]. The existing research on Voronoi diagrams on networks covers a wide range of queries including queries for the closest [12, 28], the farthest [7, 25], and the $k$-th nearest neighbors [8, 21, 31] among a finite set of sites. We refer the interested reader to, for instance, Okabe et al. [24], Okabe and Sugihara [26], Okabe and Suzuki [27], and Taniar and Rahayu [31] for guides through the vast literature on network Voronoi diagrams.

Network Voronoi diagrams relate to center problems from location analysis [12, 27]. Let $G = (V, E)$ be a network. A *center* [11] of $G$ is a vertex $v$ that minimizes the network distance to any other vertex, i.e., $\max_{u \in V} d(u, v) = \min_{x \in V} \max_{u \in V} d(u, x)$. An *absolute center* $a$ generalizes a center in that it may be placed anywhere along edges of the network, i.e., $\max_{u \in V} d(u, a) = \min_{p \in G} \max_{u \in V} d(u, p)$. A *continuous absolute center* is a point $c$ with minimal eccentricity, i.e., $\max_{p \in G} d(p, c) = \min_{q \in G} \max_{p \in G} d(p, q) = \min_{q \in G} \mathrm{ecc}(q)$.

Our results draw three ingredients from the literature on center problems: First, the absolute center [14] plays a crucial role when querying for farthest points in a tree network, as we shall see in Section 2.1. Second, the decomposition of cactus networks into *blocks*, *branches*, and *hinges* [2, 5, 17, 23] helps us exploit their tree structure [10, 16, 19] in Section 3. Third, viewing the network from different *perspectives* [2, 13, 20, 23] allows us to process branches and blocks independently, as explained in Sections 2.3 and 3.

Conversely, our constructions solve some center problems *en passant*: For example, we obtain the absolute center of a uni-cyclic network [13] as a by-product when building our data structure for queries in uni-cyclic networks. Furthermore, while it was known how to locate a single continuous absolute center in a cactus network in linear time [2], we produce the entire set of these centers in $O(n)$ time improving the $O(m^2 \log n)$ bound for general networks with $n$ vertices and $m$ edges [15].

A comprehensive summary of the literature about center problems is beyond the scope of this work. We refer readers interested in this sub-field of location analysis to a wealth of surveys [18, 22, 29, 33, 34], as well as to treatments of more recent results in the books by Kincaid [19], Shi [30], and Tansel [32].

## 1.3 Structure and Results

We introduce data structures supporting eccentricity queries and farthest-points queries for trees, cycles, uni-cyclic networks, and cactus networks. As summarized in Table 1, these data structures improve our results for general networks [4] and achieve optimal query times, sizes, and construction times.

| Type | Eccentricity Query | Farthest-Point Query | Size | Construction Time |
|---|---|---|---|---|
| Tree | $O(1)$ | $O(k)$ | $O(n)$ | $O(n)$ |
| Cycle | $O(1)$ | $O(\log n)$ | $O(n)$ | $O(n)$ |
| Uni-Cyclic | $O(\log n)$ | $O(k + \log n)$ | $O(n)$ | $O(n)$ |
| Cactus | $O(\log n)$ | $O(k + \log n)$ | $O(n)$ | $O(n)$ |
| General [4] | $O(\log n)$ | $O(k + \log n)$ | $O(m^2)$ | $O(m^2 \log n)$ |

Table 1: The traits of our data structures for queries in different types of networks, where $n$ is the number of vertices, $m$ is the number of edges, and $k$ is the number of reported farthest points.

The remainder of this work is organized as follows. In Section 2, we introduce data structures for trees,

cycles, and uni-cyclic networks. In Section 3, we construct data structures supporting eccentricity queries and farthest-point queries on cactus networks. Our approach is to reduce a cactus network to smaller networks having a sufficiently simple structure such that the data structures and query algorithms of Section 2 can be applied. In Section 4, we discuss directions for future research on closing the gap between general networks and cactus networks in Table 1.

## 2 Trees, Cycles, and Uni-Cyclic Networks

In this section, we introduce data structures for farthest-point queries in trees and cycles. We then combine these two structures to support queries in uni-cyclic networks, i.e., networks with exactly one simple cycle. These data structures have linear size and construction times while providing optimal query times, and they serve as building blocks for our data structure for cactus networks in Section 3.

### 2.1 Trees

The layout of farthest points on a tree hinges on the position of the absolute center. This point subdivides a tree into sub-trees where all points in a given sub-tree have their farthest points in other sub-trees. Conversely, each sub-tree has a set of leaves that are farthest from the absolute center and these leaves will be farthest from points in other sub-trees.

On tree networks, every farthest point is a leaf and the point $c$ whose farthest leaves are closest is an *absolute center* [11, 20]. In other words, an absolute center on a tree $T$ is a point with minimal eccentricity, i.e., $\mathrm{ecc}(c) = \min_{g \in T} \mathrm{ecc}(g)$. We say that two leaves $l$ and $l'$ of $T$ are *most distant* when they realize the maximum distance between two points on $T$, i.e., $d(l, l') = \max_{a,b \in T} d(a, b)$.

**Theorem 1** (Handler [14]). *Every tree has exactly one absolute center midway along any path connecting two most distant leaves and we can locate the absolute center of a tree with $n$ vertices in $O(n)$ time.*

Handler [14] determines the absolute center of a tree with two rounds of breadth-first-search: The first breadth-first-search starts from an arbitrary leaf $l$. With this search, we determine a farthest leaf $\bar{l}$ from $l$. We then start the second breath-first-search from $\bar{l}$ to determine a farthest leaf $\hat{l}$ from $\bar{l}$. Handler [14] shows that $\bar{l}$ and $\hat{l}$ are most distant leaves and that the absolute center is located midway on the path from $\bar{l}$ to $\hat{l}$.

We split a tree $T$ at its absolute center $c$ into sub-trees as follows. When $c$ lies on an edge $uv$ with $u \neq c \neq v$, we split $T$ into two sub-trees: the sub-tree $T_1$ containing the sub-edge $uc$, and the sub-tree $T_2$ containing the sub-edge $cv$. When $c$ lies on a vertex with neighbors $v_1, v_2, \ldots, v_r$, we split $T$ into $r$ sub-trees $T_1, T_2, \ldots, T_r$, where sub-tree $T_i$ contains the sub-edge $cv_i$. Figures 2 and 3 exemplify splitting trees at their absolute center and illustrate the following lemma that relates absolute centers with farthest points.

**Lemma 2.** *Let $T$ be a tree with absolute center $c$, let $T_i$ be one of the sub-trees obtained by splitting $T$ at $c$, and let $p$ be a point on $T_i$ with $p \neq c$. The farthest distance from $p$ in $T$ is $\mathrm{ecc}(p) = d(p, c) + \mathrm{ecc}(c)$ and the farthest points from $p$ are precisely the farthest leaves from $c$ outside of $T_i$, i.e., for every leaf $l$ we have*

$$\mathrm{ecc}(p) = d(p, l) \iff l \notin T_i \wedge \mathrm{ecc}(c) = d(c, l) \ .$$

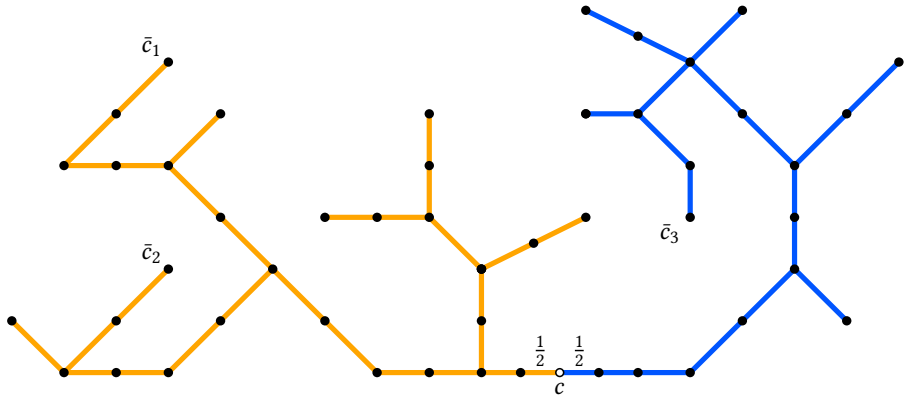

Figure 2: A tree network $T$ where all edges have unit weight unless indicated otherwise. The absolute center $c$ splits $T$ into two sub-trees $T_1$ (orange) and $T_2$ (blue). The farthest distance from $c$ is $\mathrm{ecc}(c) = 11.5$ and $c$ has three farthest leaves: $\bar{c}_1$ and $\bar{c}_2$ in $T_1$; and $\bar{c}_3$ in $T_2$. According to Lemma 2, $\bar{c}_3$ is farthest from every point on sub-tree $T_1$, whereas $\bar{c}_1$ and $\bar{c}_2$ are farthest from every point on sub-tree $T_2$.

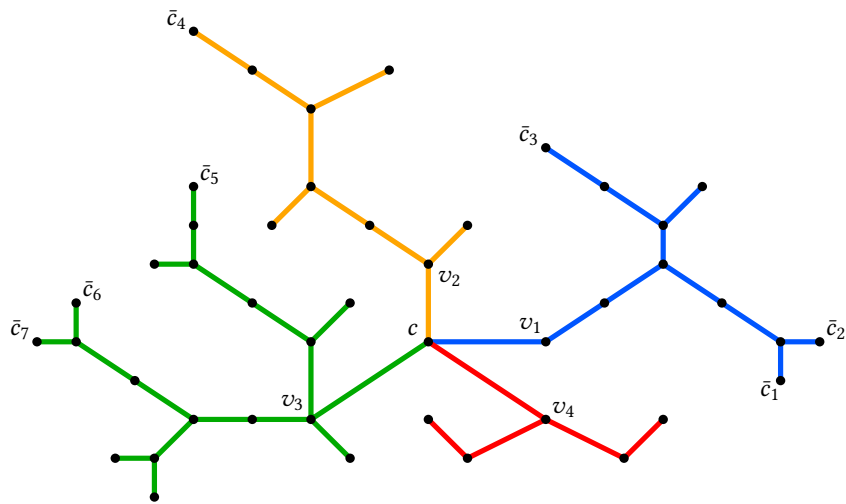

Figure 3: A tree $T$ where all edges have unit weight. The absolute center $c$ is located at a vertex and splits $T$ into four sub-trees $T_1$ (blue), $T_2$ (orange), $T_3$ (green), and $T_4$ (red). The farthest distance from $c$ is $\mathrm{ecc}(c) = 6$ and $c$ has seven farthest leaves: $\bar{c}_1, \bar{c}_2$, and $\bar{c}_3$ in $T_1$; $\bar{c}_4$ in $T_2$; $\bar{c}_5, \bar{c}_6$, and $\bar{c}_7$ in $T_3$; and none in $T_4$. Let $L_1 = \{\bar{c}_1, \bar{c}_2, \bar{c}_3\}$, $L_2 = \{\bar{c}_4\}$, $L_3 = \{\bar{c}_5, \bar{c}_6, \bar{c}_7\}$, and $L_4 = \emptyset$. According to Lemma 2, the leaves in $L_2 \cup L_3 \cup L_4$ are farthest from all points on $T_1$, the leaves in $L_1 \cup L_3 \cup L_4$ are farthest from all points on $T_2$, and the leaves in $L_1 \cup L_2 \cup L_4$ are farthest points from all points on $T_3$. All points on the red sub-tree $T_4$ share their farthest points with the absolute center $c$, since $L_4 = \emptyset$.

*Proof of Lemma 2.* We show that every path from $p$ to a farthest leaf $\bar{p}$ from $p$ passes through $c$.

Let $l$ and $\bar{l}$ be two most distant leaves of $T$. According to Theorem 1, $c$ is located midway along the path from $l$ to $\bar{l}$ with $\mathrm{ecc}(c) = d(c, l) = d(c, \bar{l})$. When splitting $T$ at $c$, the leaves $l$ and $\bar{l}$ end up in different sub-trees. Assume, without loss of generality, that $l$ and $p$ lie in different sub-trees, which implies that $c$ lies on a path from $p$ to $l$, i.e., $d(p, l) = d(p, c) + d(c, l)$.

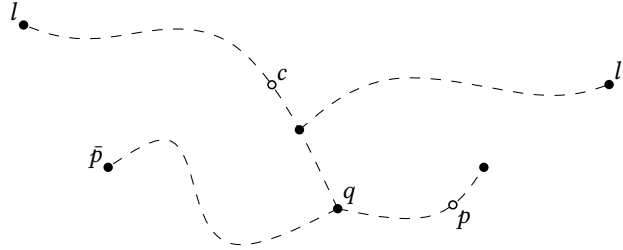

Figure 4: The impossible layout of the paths in a tree $T$ where a point $p$ on $T$ has a farthest point $\bar{p}$ such that the shortest path from $p$ to $\bar{p}$ avoids the absolute center $c$ of $T$. According to Theorem 1, the absolute center $c$ is located midway on a path between two most distant leaves $l$ and $\bar{l}$. The path from $p$ to one of these leaves—$l$ in this case—passes through $c$.

Assume, for the sake of a contradiction, that $p$ has a farthest leaf $\bar{p}$ and the path in $T$ from $p$ to $\bar{p}$ avoids $c$, i.e., $\bar{p}$ is in the same sub-tree as $p$. Let $q$ be the meeting point of the path from $c$ to $\bar{p}$ with the path from $p$ to $\bar{p}$. Figure 4 shows this (impossible) constellation in which we have

$$d(p, q) + d(q, \bar{p}) = d(p, \bar{p}) = \mathrm{ecc}(p) \geq d(p, l) = d(p, q) + d(q, l) \ ,$$

which implies $d(q, \bar{p}) \geq d(q, l)$, i.e., $l$ is no farther away from $q$ than $\bar{q}$. Since $q$ lies on the path from $p$ to $\bar{p}$ and this path avoids $c$, we have $q \neq c$ and, thus, $d(c, q) > 0$. We arrive at a contradiction via

$$d(c, \bar{p}) = d(c, q) + d(q, \bar{p}) > d(q, \bar{p}) \geq d(q, l) = d(q, c) + d(c, l) > d(c, l) \ ,$$

which implies that $\bar{p}$ is farther away from $c$ than the farthest leaf $l$, i.e., $d(c, \bar{p}) > d(c, l) = \mathrm{ecc}(c) \geq d(c, \bar{p})$. Therefore, every path from $p$ to any farthest leaf must contain the absolute center $c$ of $T$.

The farthest distance we can travel along a path from $p$ through $c$ is $d(p, c) + d(c, l) = d(p, c) + \mathrm{ecc}(c)$. Conversely, every farthest leaf from $c$ outside of $T_i$ achieves this distance and, thus, is farthest from $p$. □

We perform eccentricity queries on tree networks as follows. Let $T$ be a tree with absolute center $c$. Consider an eccentricity query from a point $p$ on edge $uv$ where $u$ is closer to $c$ than $v$, i.e., $d(c, u) < d(c, v)$. The shortest path from $p$ to $c$ leads through $u$, i.e., $d(p, c) = w_{pu} + d(u, c)$, and we have $\mathrm{ecc}(p) = d(p, c) + \mathrm{ecc}(c) = w_{pu} + d(u, c) + \mathrm{ecc}(c)$. Thus, we can determine the eccentricity of $p$ in constant time, provided that we know the eccentricity of $c$ and the network distance from $c$ to every vertex of $T$.

We perform farthest-point queries on $T$ as follows. Let $T_1, T_2, \ldots, T_r$ be the sub-trees obtained by splitting $T$ at $c$. For $i = 1, 2, \ldots, r$, let $L_i$ denote the set of farthest leaves from $c$ in sub-tree $T_i$, i.e., $L_i := \{l \in T_i \mid \mathrm{ecc}(c) = d(c, l)\}$. For a farthest point query from a point $p$ on sub-tree $T_i$ with $p \neq c$, we report all leaves in each $L_j$ with $j \neq i$ as farthest points of $p$. For a farthest-point query from the absolute center $c$, we report all leaves in $L_i$ for all $i = 1, 2, \ldots, r$. For a query point with $k$ farthest points, this takes $O(k)$ time, provided that we know the sets of farthest leaves $L_1, L_2, \ldots, L_r$ and provided that we can identify the sub-tree among $T_1, T_2, \ldots, T_r$ containing the query point in constant time.

The description of eccentricity queries and farthest-point queries on trees suggests which auxiliary data should be pre-computed. For a given tree $T$ with $n$ vertices, we first locate the absolute center $c$ of $T$ in $O(n)$ time using Handler's Algorithm [14]. Using a breadth-first-search from $c$, we perform three tasks: we compute the distances from $c$ to every vertex of $T$, we label each edge with an index indicating its sub-tree, and we determine the farthest leaves in each sub-tree, which yields the sets $L_1, L_2, \ldots, L_r$. Altogether, we spend $O(n)$ time to obtain our data structure, which is summarized in the following theorem.

**Theorem 3.** *Let $T$ be a tree network with $n$ vertices. There is a data structure with $O(n)$ construction time supporting eccentricity queries on $T$ in constant time and farthest-point queries on $T$ in $O(k)$ time, where $k$ is the number of reported farthest points.*

By storing the lengths of the lists $L_1, L_2, \ldots, L_r$, the data structure from <span style="color:red">Theorem 3</span> also allows us to count the number of farthest-points from any query point in a tree network in constant time.

## 2.2 Cycles

Let $C$ be a cycle network and let $w_C$ be the sum of all edge weights of $C$. Each point $p$ on $C$ has exactly one farthest point $\bar{p}$ located on the opposite side of $C$ with $\mathrm{ecc}(p) = d(p, \bar{p}) = w_C/2$. Supporting eccentricity queries on $C$ amounts to calculating and storing the value $w_C/2$.

To support farthest-point queries, we subdivide $C$ at each farthest point of a vertex and store a pointer from each vertex to its farthest point and vice versa. To compute this subdivision, we first locate the farthest point $\bar{v}$ for some initial vertex $v$ by walking a distance of $w_C/2$ from $v$ along $C$. As illustrated in <span style="color:red">Figure 5</span>, we then sweep a point $p$ from position $p = v$ to position $p = \bar{v}$ along $C$ while maintaining the farthest point $\bar{p}$. During this sweep we subdivide $C$ at $p$ whenever $\bar{p}$ hits a vertex and at $\bar{p}$ whenever $p$ hits a vertex. The entire sweep takes linear time, thus, the resulting data structure occupies linear space.

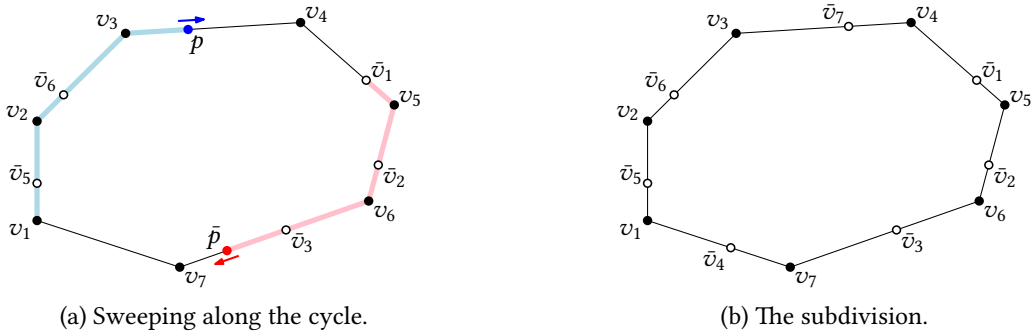

(a) Sweeping along the cycle.  (b) The subdivision.

Figure 5: A sweep along cycle $C$ starting from $p = v_1$ and the resulting subdivision of $C$. For instance, any point on sub-edge $v_5\bar{v}_2$ has its farthest point on sub-edge $\bar{v}_5 v_2$.

Using the subdivided network, we answer farthest-point queries as follows. For a query point $p$ on edge $uv$ of $C$, we first locate the sub-edge $ab$ containing $p$ using binary search. This takes $O(\log n)$ time for a cycle with $n$ vertices, since we subdivide $uv$ at most $n$ times. Let $\bar{a}$ and $\bar{b}$ be farthest from $a$ and $b$, respectively. The farthest point $\bar{p}$ from $p$ is located on sub-edge $\bar{a}\bar{b}$ at distance $w_{ap}$ from $\bar{a}$.

**Lemma 4.** *Let $C$ be a cycle network with $n$ vertices. There is a data structure with construction time $O(n)$ supporting eccentricity queries on $C$ in constant time and farthest-point queries on $C$ in $O(\log n)$ time.*

## 2.3 Uni-Cyclic Networks

A network with exactly one simple cycle is called *uni-cyclic* [13]. As illustrated in Figure 6, every uni-cyclic network $U$ consists of a cycle $C$ with trees $T_1, T_2, \ldots, T_l$ attached to $C$ at vertices $v_1, v_2, \ldots, v_l$, respectively. We refer to the trees $T_1, T_2, \ldots, T_l$ as the *branches* of $U$ and to the vertex $v_i$ as the *hinge* of branch $T_i$.

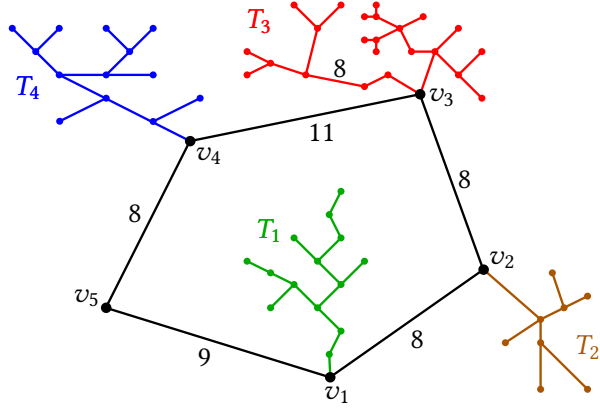

Figure 6: A uni-cyclic network with four branches. Edges have weight one whenever no weight is indicated.

Our data structure for uni-cyclic networks consists of two adjoined data structures: one for queries from the branches and one for queries from the cycle. These data structures are based on the following contractions. For each branch $T$ with hinge $v$ in a uni-cyclic network $U$, we represent all paths in $U$ starting at hinge $v$ and leading out of $T$ by an edge $vv'$ (to a dummy vertex $v'$) of weight $w_{vv'} = \text{ecc}_{U \setminus T}(v)$. We call the tree consisting of $T$ and the edge $vv'$ the *perspective* of $T$ from $U$ denoted by $\text{per}_U(T)$. For the cycle $C$, we represent the paths starting from hinge $v$ leading into a branch $T$ by an edge $vt$ (to a dummy vertex $t$) of weight $w_{vt} = \text{ecc}_T(v)$. We call the resulting network the *perspective* of $U$ from $C$ denoted by $\text{per}_U(C)$. Figure 7 summarizes the different perspectives for the network from Figure 6.

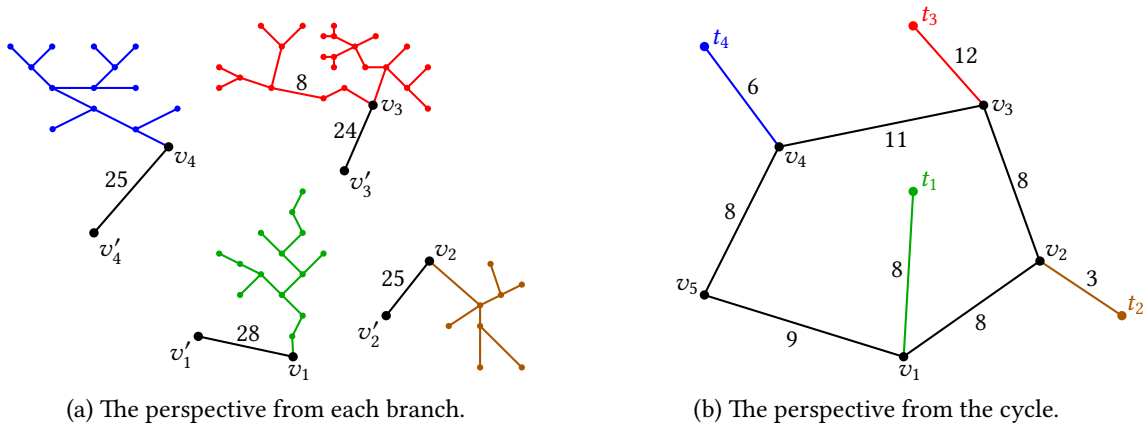

(a) The perspective from each branch.  (b) The perspective from the cycle.

Figure 7: The network from Figure 6 viewed from the branches (a) and from the cycle (b).

The data structure for queries from the branches $T_1, T_2, \ldots, T_l$ consists of the tree data structures for per($T_1$), per($T_2$), …, per($T_l$) from Section 2.1. Consider a point $p$ on branch $T_i$. An eccentricity query from $p$ in per($T_i$) yields the eccentricity of $p$ in $U$. A farthest-point query from $p$ in per($T_i$) reports the farthest points from $p$ that are located in $T_i$ and this query reports $t_i$ whenever $p$ has farthest points outside of $T_i$. As we shall explain later, we obtain the farthest points from $p$ outside of $T_i$ using a query from the hinge of $T_i$ in our data structure for the cycle perspective per($C$).

The data structure for queries from the cycle $C$ consists of two components: the first component reports the farthest points from $C$ on $C$ itself, using the data structure for cycles from Section 2.2 on $C$ ignoring the remainder of $U$. The second component reports which branches among $T_1, T_2, \ldots, T_l$, if any, contain farthest points by supporting queries for the farthest vertices among $t_1, t_2, \ldots, t_l$ from any query point on $C$ in the cycle perspective per($C$). We call this type of query a *farthest-branch query*.

### 2.3.1 Farthest-Branch Queries

We consider the perspective $\text{per}_U(C)$ for the cycle $C$ of a uni-cyclic network. The vertices $t_1, t_2, \ldots, t_l$ represent the compressed branches of $U$ in $\text{per}_U(C)$ and are connected to the hinges $v_1, v_2, \ldots, v_l$, respectively. Furthermore, let $\bar{v}_i$ denote the farthest point on $C$ from hinge $v_i$, i.e., $d_C(v_i, \bar{v}_i) = \text{ecc}_C(v_i)$.

We call a vertex $t_i$ *relevant* if there exists a point $p$ on $C$ that has $t_i$ as a farthest vertex among $t_1, t_2, \ldots, t_l$, and we call $t_i$ *irrelevant* otherwise. Knowing which of $t_1, t_2, \ldots, t_l$ are relevant will enable us to perform farthest-branch queries, i.e., report all branches containing farthest points from a query point on $C$.

**Lemma 5.** *Vertex $t_i$ is relevant if and only if $t_i$ is farthest from $\bar{v}_i$ among $t_1, t_2, \ldots, t_l$.*

*Proof.* We show both directions via indirection.

When $t_i$ is irrelevant no point on $C$—including $\bar{v}_i$—has $t_i$ as a farthest vertex among $t_1, t_2, \ldots, t_l$.

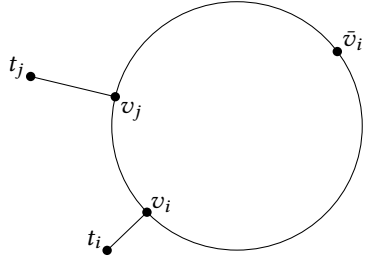

Figure 8: The constellation of $t_i$ and $t_j$ where $v_j$ lies on the clockwise path from $v_i$ to $\bar{v}_i$.

Conversely, let there be some vertex $t_j$ that is farther away from $\bar{v}_i$ than $t_i$, i.e., $d(\bar{v}_i, t_i) < d(\bar{v}_i, t_j)$. As illustrated in Figure 8, $v_j$ lies either on the clockwise or counter-clockwise path from $t_i$ to $\bar{v}_i$. Since either path is a shortest path, we have $d(t_i, \bar{v}_i) = d(t_i, v_j) + d(v_j, \bar{v}_i)$. Thus, $t_j$ is farther from $v_j$ than $t_i$, as

$$d(t_i, v_j) = d(t_i, \bar{v}_i) - d(\bar{v}_i, v_j) < d(t_j, \bar{v}_i) - d(\bar{v}_i, v_j) = d(t_j, v_j) \ .$$

Now $t_i$ is irrelevant, because $t_j$ is farther away from any point $p$ on $C$ than $t_i$, since

$$d(p, t_i) \leq d(p, v_j) + d(v_j, t_i) < d(p, v_j) + d(v_j, t_j) = d(p, t_j) \ . \qquad \square$$

According to Lemma 5, a vertex $t_i$ is irrelevant when there is some other vertex $t_j$ with $d(t_i, \bar{v}_i) < d(t_j, \bar{v}_i)$. In this case, we say that $t_i$ *is dominated by* $t_j$ and write $t_i \prec t_j$. Algorithm 1 below computes the relevant vertices using dominance. We begin with a circular list containing the vertices $t_1, t_2, \ldots, t_l$ in the order as the hinges $v_1, v_2, \ldots, v_l$ appear along the cycle $C$. We traverse the list in counterclockwise order and delete vertices whenever they are dominated by their successor (succ) or their predecessor (pred). We mark a vertex $t$ as processed if we can neither remove $t$ nor any of its neighbors based on this criteria.

---

**Algorithm 1:** Determining the relevant vertices

    **input**   : The vertices $t_1, t_2, \ldots, t_l$ stored in a circular list.
    **output** : The relevant vertices among $t_1, t_2, \ldots, t_l$.

1   Mark each $t_1, t_2, \ldots, t_l$ as unprocessed;
2   $t \leftarrow t_1$;
3   **while** $t$ *is unprocessed* **do**
4      |   **if** $t \prec \text{PRED}(t)$ **or** $t \prec \text{SUCC}(t)$ **then**
5      |      |   $t \leftarrow \text{SUCC}(t)$;
6      |      |   $\text{DELETE}(\text{PRED}(t))$;
7      |   **else if** $\text{PRED}(t) \prec t$ **then** $\text{DELETE}(\text{PRED}(t))$;
8      |   **else if** $\text{SUCC}(t) \prec t$ **then** $\text{DELETE}(\text{SUCC}(t))$;
9      |   **else** /* $t \not\prec \text{PRED}(t) \not\prec t \not\prec \text{SUCC}(t) \not\prec t$ */
10     |      |   Mark $t$ as processed;
11     |      |   $t \leftarrow \text{SUCC}(t)$;
12     |   **end**
13   **end**

---

**Invariant 6.** *The following invariants hold whenever Algorithm 1 marks a vertex as processed in Line 10.*

   *(i) Every marked vertex $t$ dominates none of its current neighbors, i.e., $\text{PRED}(t) \not\prec t$ and $\text{SUCC}(t) \not\prec t$.*

   *(ii) Every marked vertex $t$ is dominated by none of its current neighbors, i.e., $t \not\prec \text{PRED}(t)$ and $t \not\prec \text{SUCC}(t)$.*

*Proof.* Assume, for the sake of a contradiction, that there is a marked vertex $t_i$ with a neighbor $t_j$ such that $t_i \prec t_j$ or $t_j \prec t_i$ when Line 10 is executed. Assume, without loss of generality, that $t_j$ has been marked after $t_i$ or that $t_j$ has never been marked so far. When we marked $t_i$ as processed, $t_i$ and $t_j$ were not neighbors. When $t_j$ became a neighbor of $t_i$, the previous neighbor $t_k$ of $t_i$ was either deleted in Line 6 with $t_j$ assuming the role of variable $t$ in Line 5 or $t_k$ was deleted in Lines 8 with $t = t_j$. In both cases variable $t$ would be at $t_j$ with $t_i$ as direct neighbor and we would have deleted $t_i$ in Line 7 or in Line 8 before marking any other vertex as processed. Therefore, we have never marked $t_i$, which contradicts our assumption. □

As a consequence of Invariant 6, no vertex dominates any neighbor when Algorithm 1 terminates.

We are now ready to prove the correctness of Algorithm 1. We pre-compute the distances from $v_1$ to all other vertices along $C$ while constructing per($C$). This allows us to compare distances along $C$ in constant time and, thus, enables us to determine in constant time whether one vertex dominates another.

**Theorem 7.** *Let $S$ be the perspective of a uni-cyclic network $U$ from its cycle $C$, and let $t_1, t_2, \ldots, t_l$ be the vertices of $S$ representing the $l$ branches of $U$ given in clockwise order. Algorithm 1 computes all relevant vertices among $t_1, t_2, \ldots, t_l$ in $O(l)$ time, provided that checking for dominance takes constant time.*

*Proof.* In each iteration of the while-loop of Algorithm 1, we either delete some vertex or we mark the vertex stored in $t$ as processed ensuring that it will never assume the role of $t$ again. Therefore, Algorithm 1 terminates in $O(l)$ steps, provided that checking for dominance is a constant time operation.

Algorithm 1 never deletes a relevant vertex, since we only delete vertices that are dominated and, thus, irrelevant. Consider the circular list of those vertices that remain after Algorithm 1 terminates. We assume, for the sake of a contradiction, that this list contains irrelevant vertices. Let $t$ be a relevant vertex and let $t_{\mathrm{cw}}$ be the first irrelevant vertex in clockwise direction from $t$, and let $t_{\mathrm{ccw}}$ be the first irrelevant vertex in counterclockwise direction from $t$. Since no vertex in the final list dominates any of its neighbors by Invariant 6, the final list contains at least four vertices with at least one vertex between $t$ and $t_{\mathrm{cw}}$ and at least one vertex between $t$ and $t_{\mathrm{ccw}}$. The argumentation below remains valid when $t_{\mathrm{cw}}$ and $t_{\mathrm{ccw}}$ coincide.

Let $\bar{v}$, $\bar{v}_{\mathrm{cw}}$ and $\bar{v}_{\mathrm{ccw}}$ be the farthest point on $C$ from $t$, $t_{\mathrm{cw}}$, and $t_{\mathrm{ccw}}$, respectively. We distinguish the two cases illustrated in Figure 9, based on the relative positions of $t$, $\bar{v}$, and $\bar{v}_{\mathrm{ccw}}$.

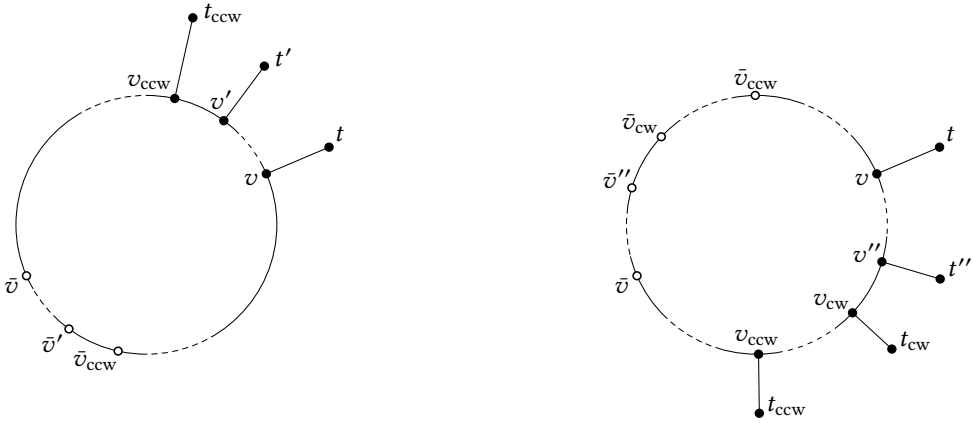

(a) The point $\bar{v}_{\mathrm{ccw}}$ lies clockwise between $t$ to $\bar{v}$.　　(b) The point $\bar{v}_{\mathrm{ccw}}$ lies counterclockwise between $t$ to $\bar{v}$.

Figure 9: The cyclic order of the branch representing vertices $t$, $t_{\mathrm{ccw}}$, $t_{\mathrm{cw}}$, $t'$, $t''$, and their corresponding antipodal points in the two cases from the proof of Theorem 7.

Consider the case when the point $\bar{v}_{\mathrm{ccw}}$ lies on the clockwise path from $t$ to $\bar{v}$, as illustrated in Figure 9a. Let $t' := \mathrm{PRED}(t_{\mathrm{ccw}})$ be the clockwise neighbor of $t_{\mathrm{ccw}}$ and let $\bar{v}'$ be the farthest point from $t_{\mathrm{ccw}}$ on $C$. The vertices $\bar{v}_{\mathrm{ccw}}$, $\bar{v}'$, and $\bar{v}$ appear clockwise in this order along $C$, which implies that $\bar{v}_{\mathrm{ccw}}$ lies on a shortest path from $t$ to $\bar{v}'$, i.e., $d(t, \bar{v}') = d(t, \bar{v}_{\mathrm{ccw}}) + d(\bar{v}_{\mathrm{ccw}}, \bar{v}')$. Furthermore, we have $t_{\mathrm{ccw}} \prec t$ and $t' \not\prec t$, since $t_{\mathrm{ccw}}$ was the first dominated vertex in clockwise direction from $t$. Together, this yields

$$d(t_{\mathrm{ccw}}, \bar{v}_{\mathrm{ccw}}) \overset{t_{\mathrm{ccw}} \prec t}{<} d(t, \bar{v}_{\mathrm{ccw}}) = d(t, \bar{v}') - d(\bar{v}', \bar{v}_{\mathrm{ccw}}) \overset{t' \not\prec t}{\leq} d(t', \bar{v}') - d(\bar{v}', \bar{v}_{\mathrm{ccw}}) = d(t', \bar{v}_{\mathrm{ccw}}) \ ,$$

which implies $d(t_{\mathrm{ccw}}, \bar{v}_{\mathrm{ccw}}) < d(t', \bar{v}_{\mathrm{ccw}})$, i.e., $t_{\mathrm{ccw}} \prec t'$. This contradicts Invariant 6.

Consider the case when $\bar{v}_{\mathrm{ccw}}$ lies on the counterclockwise path from $t$ to $\bar{v}$, as illustrated in Figure 9b. This implies that the vertices $t$, $t_{\mathrm{cw}}$, $t_{\mathrm{ccw}}$, and $\bar{v}$ appear clockwise in this order (with potentially $t_{\mathrm{cw}} = t_{\mathrm{ccw}}$). Therefore, $\bar{v}_{\mathrm{cw}}$ also lies on the counterclockwise path from $t$ to $\bar{v}$. Similarly, to the above, we derive the contradiction $t_{\mathrm{cw}} \prec t''$ where $t'' := \mathrm{SUCC}(t_{\mathrm{cw}})$ is the counterclockwise neighbor of $t_{\mathrm{cw}}$. Therefore, $t_{\mathrm{cw}}$ (and thus $t_{\mathrm{ccw}}$) cannot exist and there are no irrelevant vertices in the circular list produced by Algorithm 1. □

We have a circular list of the relevant vertices among $t_1, t_2, \ldots, t_l$. Using this list, we pre-process $S$ to support farthest-branch queries as follows. We pick any relevant vertex $t_i$ and traverse $C$ counterclockwise starting from $\bar{v}_i$ keeping track of the farthest branch. Since $t_i$ and its counterclockwise successor $t_j = \text{succ}(t_i)$ are relevant, $\bar{v}_i$ has $t_i$ as its farthest branch and $\bar{v}_j$ has $t_j$ as its farthest branch. At some point $p_{ij}$ between $\bar{v}_i$ and $\bar{v}_j$ the farthest branch changes from $t_i$ to $t_j$; we subdivide $C$ at $p_{ij}$ and store $t_i$ as farthest branch for the (sub)edges from $\bar{v}_i$ to $p_{ij}$. We continue subdividing $C$ in this fashion into at most $l$ chains, one for each relevant vertex. Using a binary search on these chains, we answer farthest-branch queries in $O(\log l)$ time. Figure 10 illustrates an example of this subdivision for the network from Figure 6. As in all figures in this section, edges have unit weight unless indicated otherwise.

**Theorem 8.** *Let $U$ be a uni-cyclic network with $n$ vertices and cycle $C$ of length $l$. There is a data structure with construction time $O(n)$ supporting farthest-branch queries from any query point on $C$ in $O(\log l)$ time.*

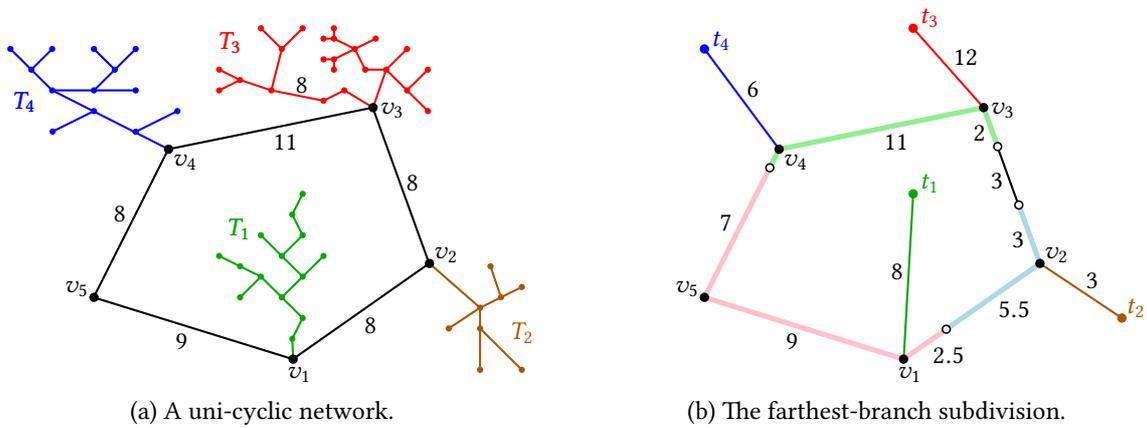

(a) A uni-cyclic network.  (b) The farthest-branch subdivision.

Figure 10: The farthest-branch subdivision (b) of the uni-cyclic network from (a). The sub-edges in (b) are shaded in a colour matching their farthest branch. On black sub-edges, no branch contains farhest points—instead, the farthest point lies on the opposite side of the cycle. In this example, no points on the cycle have farthest points in branch $T_2$, i.e., $t_2$ is irrelevant.

### 2.3.2 Queries in Uni-Cyclic Networks

We perform an eccentricity query from a point $q$ on $U$ as follows. When $q$ lies on some branch $T$, we use the tree data structures for the perspective $T' = \text{per}(T)$ from $T$, as $\text{ecc}_{T'}(q) = \text{ecc}_U(q)$ by construction of $T'$. When $q$ lies on the cycle $C$, we first compute the distance $d_B := \max_{i=1}^{l} d(q, t_i)$ from $q$ to the farthest vertex among $t_1, t_2, \ldots, t_l$ in the perspective $S = \text{per}(C)$ from $C$, using our data structure for farthest-branch queries in $S$. Then, we compute the farthest distance $\text{ecc}_C(q)$ from $q$ on $C$, using the cycle data structure for $C$. The greater distance of the two is the eccentricity of $q$ in $S$ and, thus, the eccentricity of $q$ in $U$, i.e., $\text{ecc}_U(q) = \text{ecc}_S(q) = \max\{d_B, \text{ecc}_C(q)\}$. This way, we answer eccentricity queries from branches in constant time and eccentricity queries from the cycle in $O(\log l)$ time.

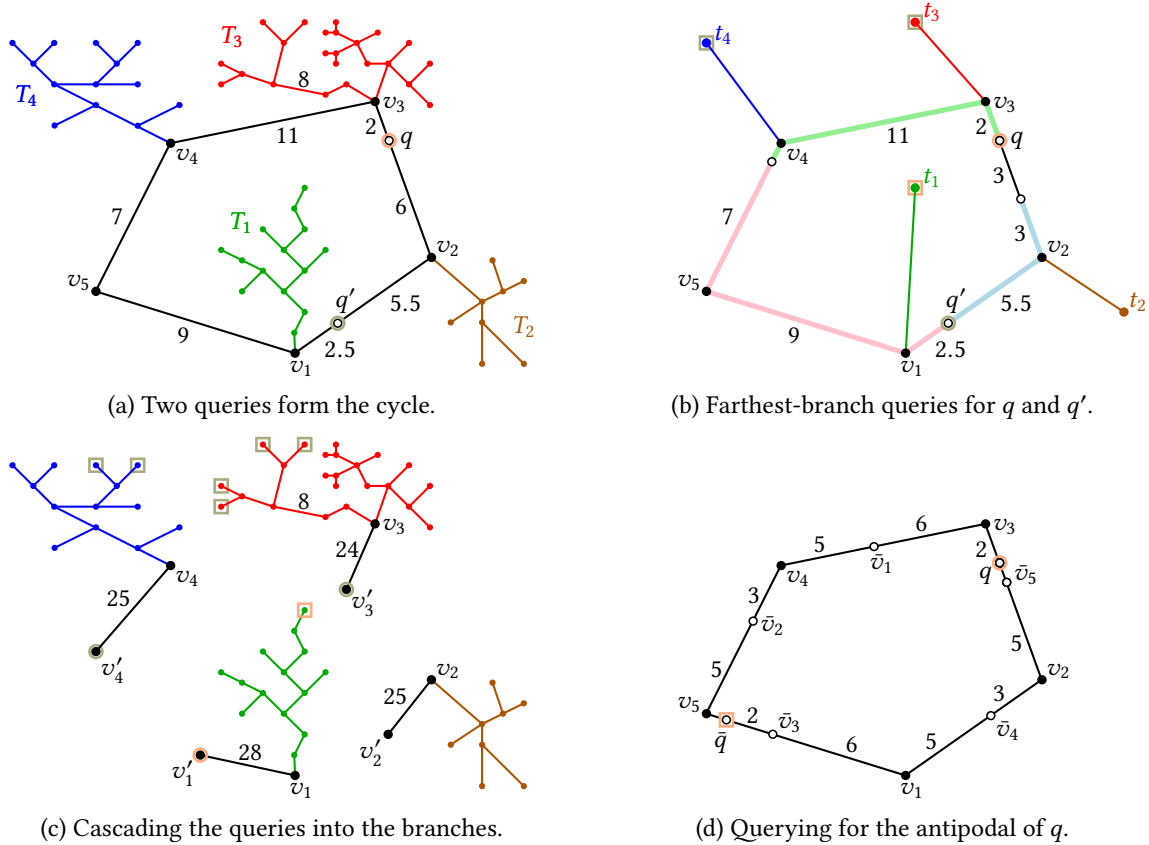

(a) Two queries form the cycle.

(b) Farthest-branch queries for $q$ and $q'$.

(c) Cascading the queries into the branches.

(d) Querying for the antipodal of $q$.

Figure 11: The farthest-point queries from the points $q$ (salmon) and $q'$ (ocker) on the cycle of a uni-cyclic network (a). We perform a farthest-branch query for both query points (b) and then cascade the queries into the perspectives from the branches (c) and into the cycle (d) as needed.

We perform a farthest-point query from a point $q$ on the cycle $C$ of $U$ as illustrated in Figure 11. We first perform a farthest branch query from $q$ in $S$ and a farthest point query from $q$ in $C$. We report the antipodal of $q$ on $C$ if it is farthest from $q$ in $U$ and then cascade the query in each branch. More precisely, if $T$ was a branch containing farthest points from $q$ then perform a farthest-point query from $v'$ in the perspective $T' = \text{per}_U(T)$ of $U$ from $T$, where $v'$ is the vertex representing the exterior of $T$. The farthest leaves from $v'$ in $T'$ are also farthest from any point outside of $T$ that has farthest points in $T$.

We perform a farthest-point query from a point $q$ on a branch $T$ of $U$ with hinge $v$ as follows. We begin with a farthest point query in the perspective $T' = \text{per}(T)$ of $U$ from $T$. This reports all farthest points from $q$ in $T$ and, potentially, the point $v'$ representing the exterior of $T$. When cascading the query from $T'$ into the perspective $S = \text{per}(C)$ of $U$ from the cycle $C$, we need to perform a farthest-point query in $S$ from the vertex $t$ representing $T$. However, we have no structure to support this query directly. Instead, we query from $v$ in $S$, which leads to a *good case* and a *bad case*.

In the good case, shown in Figure 12, $v$ has some farthest point in $S$ other than $t$, which means that the farthest points from $t$ are the farthest points from $v$ (potentially excluding $t$ itself). In the bad case, shown in Figure 13, $t$ is the only farthest point from $v$ in $S$. Fortunately, the bad case can only appear for exactly one branch of $U$, as $t$ being the only farthest vertex from $v$ implies that $t$ is farthest from all other vertices of $S$ as well. Therefore, we can deal with the bad case by computing the farthest points from $v$ in the network $S - vt$ (during the preprocessing phase) and storing the result with $v$. We use this only when cascading a query from $T$ into $S$, i.e., when we know that there are farthest points from the query point $q$ outside of $T$.

The following theorem summarizes our data structure for uni-cyclic networks.

**Theorem 9.** *Let $U$ be a uni-cyclic network with $n$ vertices and a cycle of length $l$. There is a data structure with size and construction time $O(n)$ supporting eccentricity queries from the branches of $U$ in $O(1)$ time, eccentricity queries from the cycle in $O(\log l)$ time, farthest-point queries from branches in $O(k)$ time, and farthest-point queries from the cycle in $O(k + \log l)$ time, where $k$ is the number of reported farthest points.*

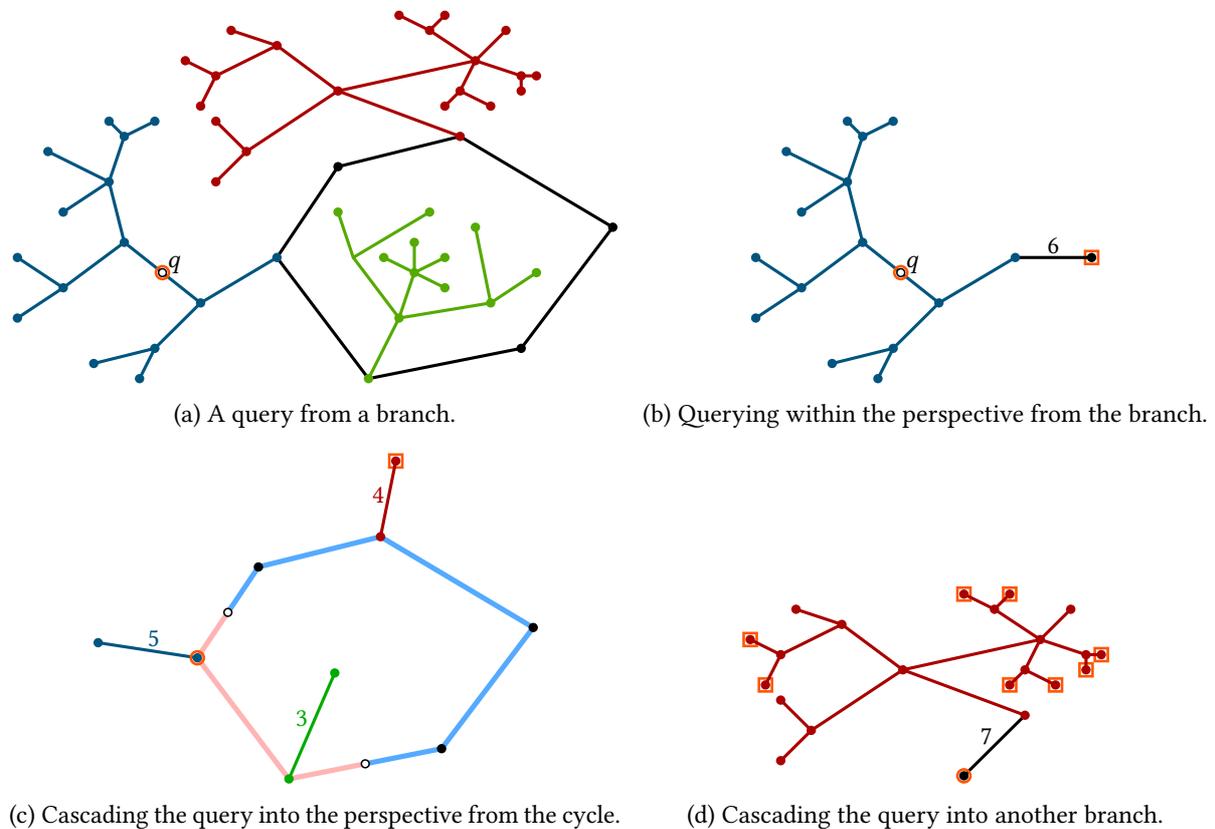

(a) A query from a branch.

(b) Querying within the perspective from the branch.

(c) Cascading the query into the perspective from the cycle.

(d) Cascading the query into another branch.

Figure 12: A farthest-point query from a branch of a unicyclic network (a). We perform a farthest-point query in the perspective from the branch containing the query point (b). We cascade the query into the perspective from the cycle (c) and then into another branch as needed (d).

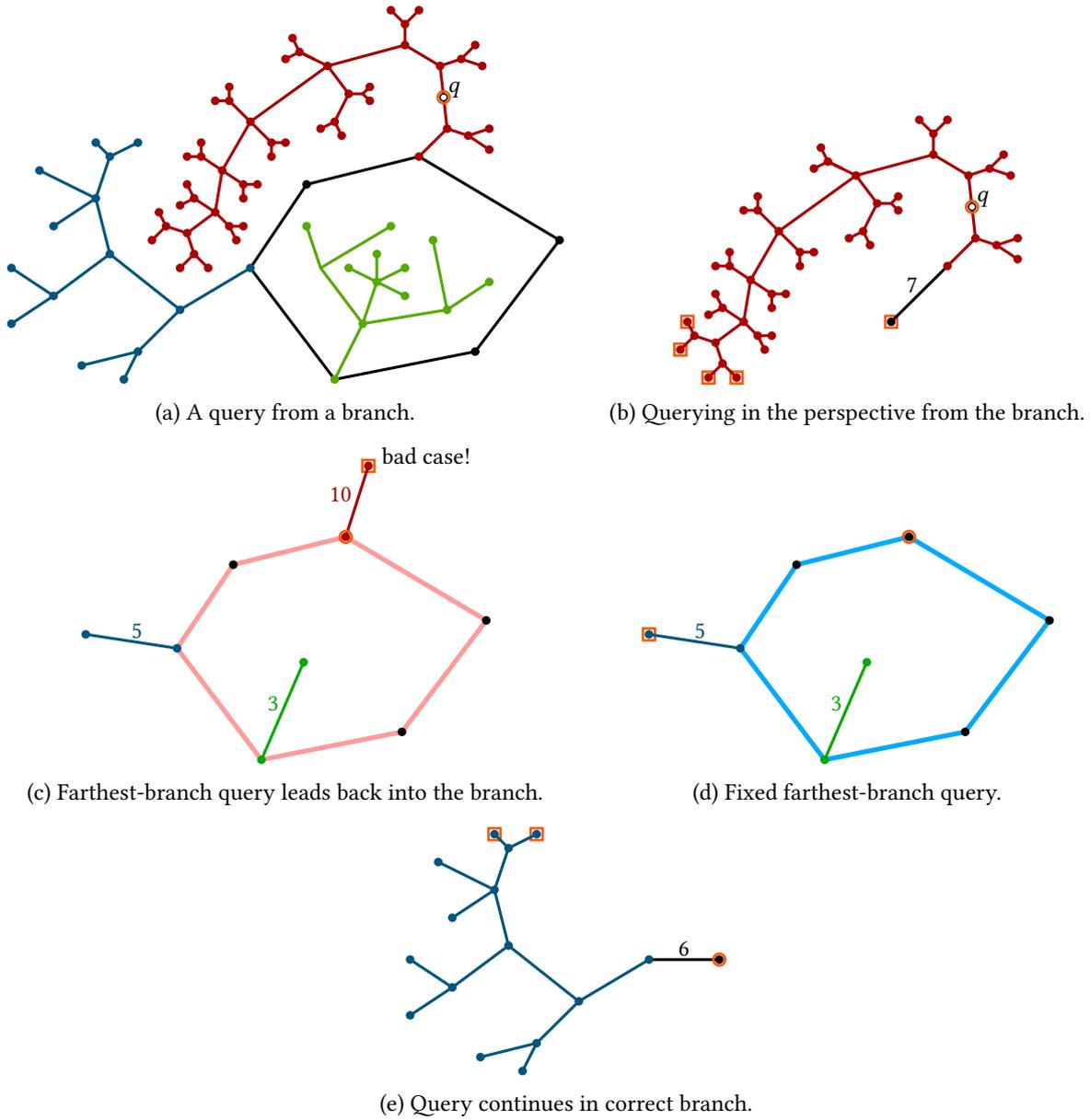

(a) A query from a branch.

(b) Querying in the perspective from the branch.

(c) Farthest-branch query leads back into the branch.

(d) Fixed farthest-branch query.

(e) Query continues in correct branch.

Figure 13: The bad case for a farthest-point query from a branch $T$ (red) of a unicyclic network (a). We perform a farthest-point query in the perspective from the branch containing the query point (b). We cascade the query into the perspective from the cycle (c), however, this query sends us back into the red branch, since the red branch is farthest on the entire cycle. To fix this, we remove the edge representing the red branch (d) and repeat the query in the resulting network. This time, we obtain the correct branch where we cascade the query (e) to determine the farthest points in this (blue) branch. Observe that the fixed network (d) is irrelevant for queries from other branches which will use the normal perspective from the cycle (c).

## 3 Cactus Networks

In this section, we construct a data structure supporting eccentricity queries and farthest-point queries on cactus networks. Recall that a cactus networks is a network in which no two simple cycles share an edge.

The following notions prove useful when describing cactus networks; examples are shown in Figure 14. A *cut-vertex* is a vertex whose removal increases the number of connected components, a *block* is a maximal connected sub-graph without cut-vertices and with at least three vertices, and a *branch* is a (maximal) tree that remains when removing the edges of all blocks and all resulting isolated vertices. We treat blocks and branches in a similar fashion, so we refer to a sub-network $B$ as a *bag* when $B$ is either a block or a branch. Decomposing a network with $n$ vertices into its blocks and branches takes $O(n)$ time [17]. A *hinge* is a vertex contained in more than one bag. The *tree structure* of a cactus network $G$, denoted by $T(G)$, is the tree whose vertices are the hinges and bags of $G$ where a hinge $h$ is connected to a bag $B$ when $h$ lies in $B$ [16].

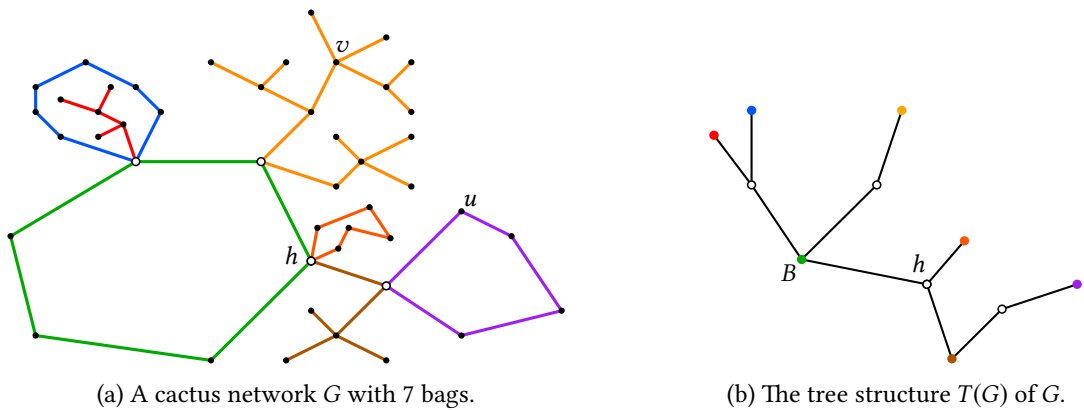

(a) A cactus network $G$ with 7 bags.

(b) The tree structure $T(G)$ of $G$.

Figure 14: A cactus network (a) together with its tree structure (b). The blocks and branches are indicated in colours and the hinges connecting these bags are marked as empty discs. For example, the green edges form a block and the red edges form a branch. Vertex $h$ is a hinge, because $h$ is a cut-vertex contained in more than one bag; vertex $v$ is a cut-vertex but not a hinge, since $v$ is only contained in the yellow branch; and vertex $u$ is not a cut-vertex and, thus, also not a hinge.

Let $B$ be a bag containing hinge $h$. The component containing $h$ after removing all edges of $B$ is called the *bag-cut* of $B$ at $h$, denoted by bcut($B, h$). The component containing $h$ after removing all edges of bcut($B, h$) is called the *co-bag-cut* of $B$ at $h$, denoted by co-bcut($B, h$). Figure 15 gives examples of bag-cuts and co-bag-cuts for the cactus network from Figure 14.

We view an (undirected) edge from $B$ to $h$ in the tree structure as two (directed) arcs $B \rightarrow h$ and $h \rightarrow B$. As illustrated in Figures 15c and 15d, we associate bcut($B, h$) with $B \rightarrow h$ and we associate co-bcut($B, h$) with $h \rightarrow B$. We use this correspondence as orientation. For example, we store the eccentricity of $h$ in bcut($B, h$) with the arc $B \rightarrow h$ and the eccentricity of $h$ in co-bcut($B, h$) with the arc $h \rightarrow B$.

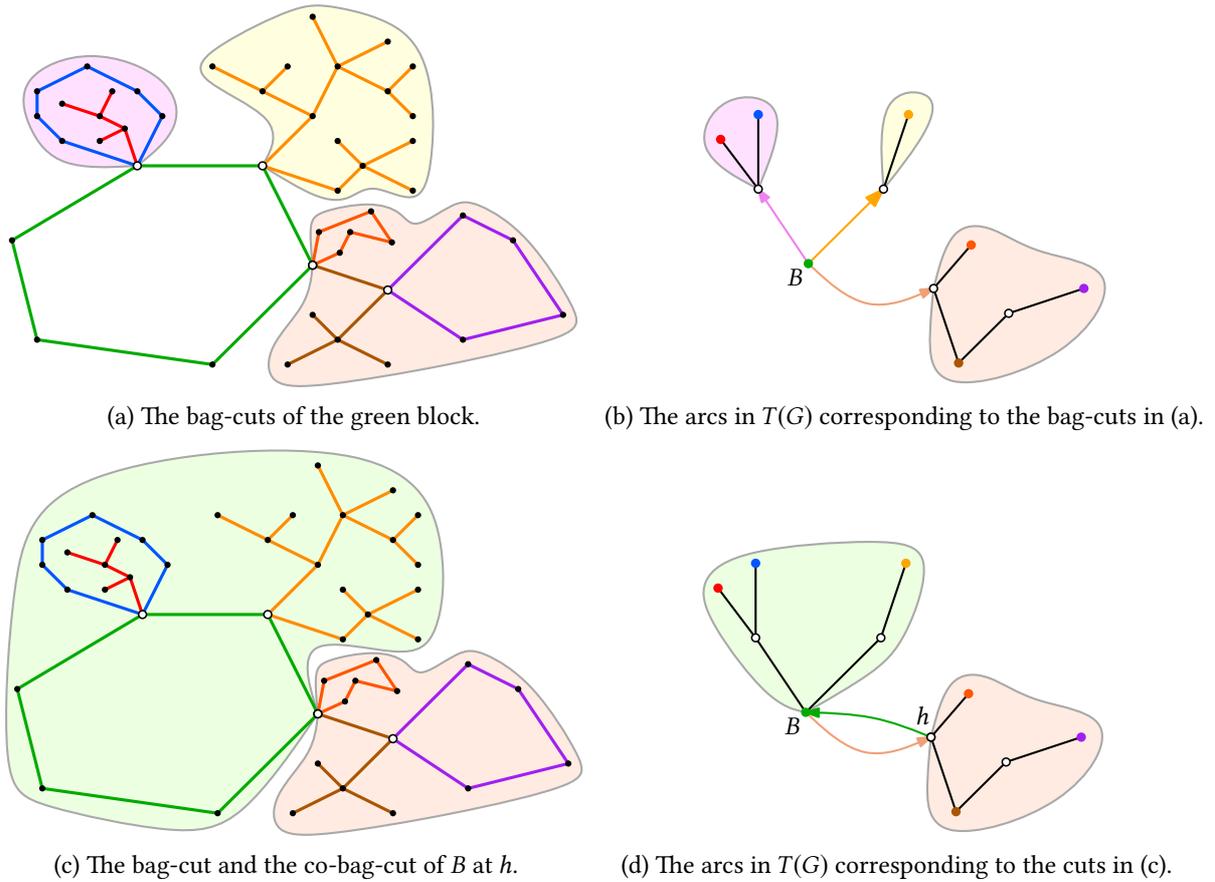

(a) The bag-cuts of the green block.

(b) The arcs in $T(G)$ corresponding to the bag-cuts in (a).

(c) The bag-cut and the co-bag-cut of $B$ at $h$.

(d) The arcs in $T(G)$ corresponding to the cuts in (c).

Figure 15: The bag-cuts of the green block (a) together with their corresponding arcs (of matching colour) in the tree structure (b). The bag-cut (salmon) and co-bag-cut (light green) corresponding to the two arcs of an edge from $B$ to $h$ in the tree structure (d) and in the network itself (c).

## 3.1 Eccentricity Queries

To support eccentricity queries on a bag $B$ of a network $G$, we compress the bag-cuts of $B$ like we compress the branches of uni-cyclic networks: for any hinge $h \in B$ we replace bcut$(B, h)$ with a vertex $\hat{h}$ and an edge $h\hat{h}$ whose weight is the largest distance from $h$ to any point in bcut$(B, h)$, i.e., $w_{h\hat{h}} = \text{ecc}_{\text{bcut}(B,h)}(h)$. We refer to the resulting network as the *perspective* of $G$ from $B$, denoted by per$_G(B)$. The perspective of $G$ from bag $B$ preserves farthest distances of $G$, i.e., we have $\text{ecc}_{\text{per}_G(B)}(p) = \text{ecc}_G(p)$ for all $p$ on $B$.

The perspective from a branch is a tree and the perspective from a block of a cactus network is a uni-cyclic network. Since, we already have efficient data structures for trees and uni-cyclic networks, the challenge lies in constructing these perspectives in linear time. As illustrated in [Figure 16](#), we first construct the perspective per$(B^*)$ of an arbitrary bag $B^*$ by computing the extended shortest path tree of each hinge $h^*$ of $B^*$ in bcut$(B^*, h^*)$. This takes linear time using a modified breadth-first search from each hinge $h^*$ where we cut each simple cycle $C$ at the antipodal $\bar{v}$ of the vertex $v$ where we first enter $C$. Using the data structure for queries in per$(B^*)$ and the extended shortest path trees from the hinges of $B^*$, we then construct the perspectives from the bags adjacent to $B^*$ and continue in a breadth-first search fashion.

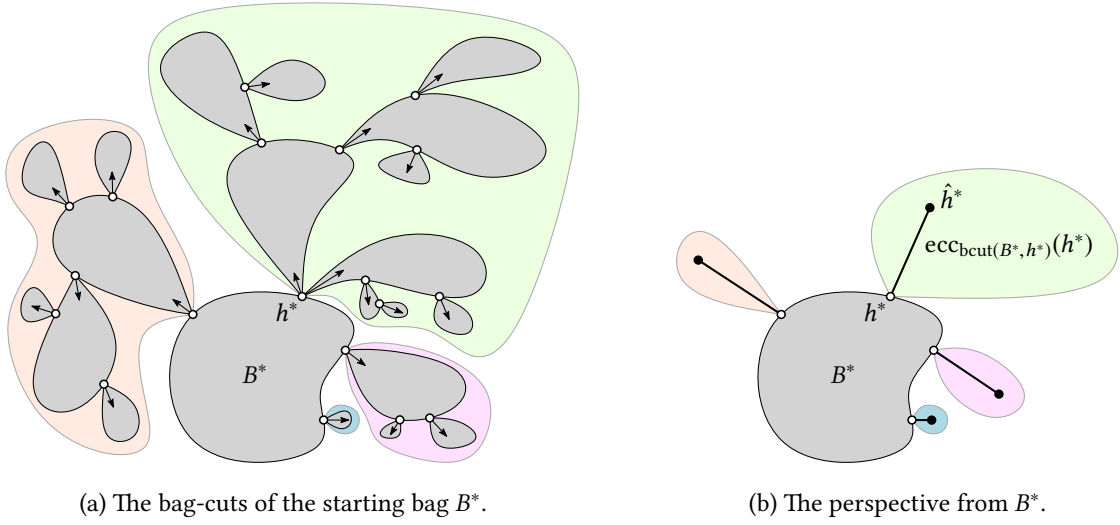

(a) The bag-cuts of the starting bag $B^*$.

(b) The perspective from $B^*$.

Figure 16: A schematic view of the construction of the perspective $\mathrm{per}(B^*)$ from the initial bag $B^*$. For every hinge $h^*$ of $B^*$, the bag-cut $\mathrm{bcut}(B^*, h^*)$ is replaced with an edge $h^*\hat{h}^*$ weighted with the largest distance from $h^*$ into $\mathrm{bcut}(B^*, h^*)$. The extended shortest path trees from the hinges of $B^*$ are outlined as arrows from the hinges of the network. An arrow pointing into the co-bag-cut of bag $B$ at hinge $h$ means that we obtain the extended shortest path tree from $h$ into co-$\mathrm{bcut}(B, h)$ as a by-product of the construction of the perspective from $B^*$.

Let $B'$ be a bag neighboring $B^*$ at hinge $h^*$, as shown in Figure 17. For all hinges $h'$ of $B'$ with $h' \neq h^*$, the extended shortest path tree of $h'$ in $\mathrm{bcut}(B', h')$ is a sub-tree of the extended shortest path tree of $h^*$ in $\mathrm{bcut}(B^*, h^*)$. So we know the weight of the edge $\hat{h}'h'$ in $\mathrm{per}(B')$, where $\hat{h}'$ represents $\mathrm{bcut}(B', h')$. At hinge $h^*$, we have the extended shortest path trees of $h^*$ in every co-bag-cut at $h^*$ except for co-$\mathrm{bcut}(B^*, h^*)$, the one leading back into $B^*$. The eccentricity of $h^*$ in $\mathrm{bcut}(B', h^*)$ is the largest distance from $h^*$ into any co-bag-cut co-$\mathrm{bcut}(B, h^*)$ for all bags $B$ neighboring $B'$ at $h^*$, i.e.,

$$\mathrm{ecc}_{\mathrm{bcut}(B', h^*)}(h^*) = \max\{\mathrm{ecc}_{\text{co-}\mathrm{bcut}(B, h^*)}(h^*) \mid B \text{ is a bag with } h^* \in B \neq B'\} \ . \tag{1}$$

We need $\mathrm{ecc}_{\text{co-}\mathrm{bcut}(B^*, h^*)}(h^*)$ to compute $\mathrm{per}(B')$. The difference between the perspective $\mathrm{per}_G(B^*)$ of $G$ from $B^*$ and the perspective $\mathrm{per}_{\text{co-}\mathrm{bcut}(B^*, h^*)}(B^*)$ of co-$\mathrm{bcut}(B^*, h^*)$ from $B^*$ is that the latter lacks the edge $\hat{h}^*h^*$ where $\hat{h}^*$ represents $\mathrm{bcut}(B^*, h^*)$ in $\mathrm{per}_G(B^*)$. We avoid constructing $\mathrm{per}_{\text{co-}\mathrm{bcut}(B^*, h^*)}(B^*)$ using the following observation. The farthest points from $\hat{h}^*$ in $\mathrm{per}_G(B^*, h^*)$ are also farthest from $h^*$ in $\mathrm{per}_{\text{co-}\mathrm{bcut}(B^*, h^*)}(B^*) = \mathrm{per}_G(B^*, h^*) - \hat{h}^*h^*$, but they are closer by $w_{h^*\hat{h}^*}$, i.e.,

$$\mathrm{ecc}_{\text{co-}\mathrm{bcut}(B^*, h^*)}(h^*) = \mathrm{ecc}_{\mathrm{per}(B^*) - \hat{h}^*h^*}(h^*) = \mathrm{ecc}_{\mathrm{per}(B^*)}(\hat{h}^*) - w_{h^*\hat{h}^*} = \mathrm{ecc}_{\mathrm{per}(B^*)}(\hat{h}^*) - \mathrm{ecc}_{\mathrm{bcut}(B^*, h^*)}(h^*) \ .$$

To see this, consider the extended shortest path tree from $\hat{h}^*$ in $\mathrm{per}_G(B^*)$. Removing the edge $\hat{h}^*h^*$ from this tree yields the extended shortest path tree from $h^*$ in $\mathrm{per}_G(B^*) - \hat{h}^*h^* = \text{co-}\mathrm{bcut}(B^*, h^*)$. In this way, we obtain $\mathrm{ecc}_{\text{co-}\mathrm{bcut}(B^*, h^*)}(h^*)$ with a constant time query from $\hat{h}^*$ in $\mathrm{per}_G(B^*)$, as illustrated in Figure 17d.

Finally, we obtain the missing value $\mathrm{ecc}_{\mathrm{bcut}(B', h^*)}(h^*)$ by taking the maximum in (1). However, we need to avoid a dependence on the number of bags containing $h^*$, i.e., the degree of $h^*$ in the tree structure.

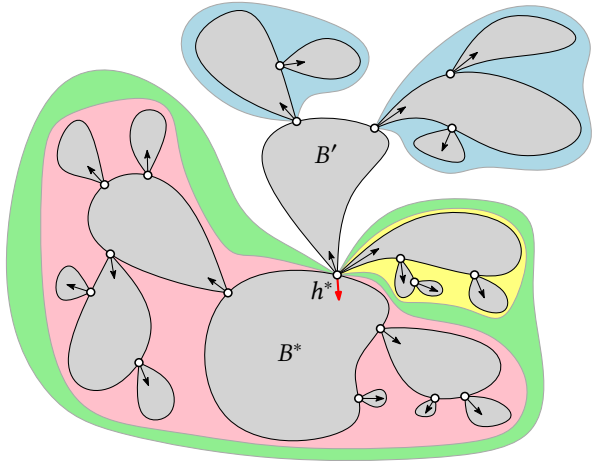

(a) The bag-cuts of neighboring bag $B'$.

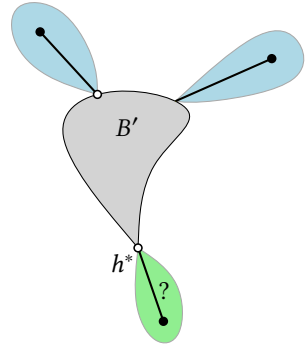

(b) The missing weight in the perspective from $B'$.

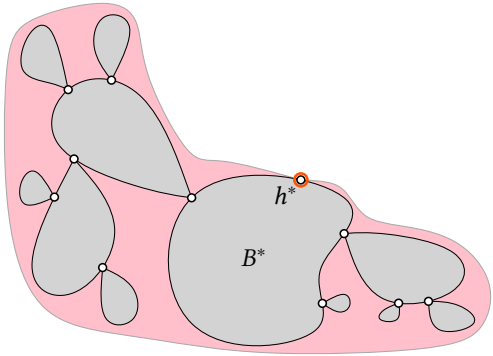

(c) The desired query in co-bcut$(B^*, h^*)$.

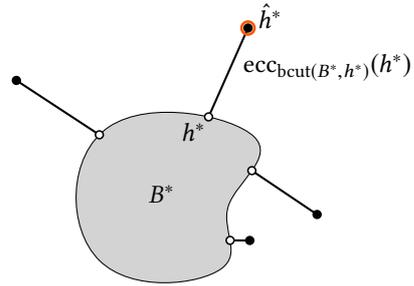

(d) The actual query in per$(B^*)$.

Figure 17: A schematic view of the construction of the perspective of a bag $B'$ neighboring the initial bag $B^*$. From the construction of per$(B^*)$, we already have the distance information for the bag-cuts (blue) of $B'$ at all hinges except for the one at $h^*$ (green) and we have the distances for all co-bag-cuts of neighbors of $B'$ (yellow) other than co-bcut$(B^*, h^*)$ (red). To obtain the missing value $\mathrm{ecc}_{\mathrm{co\text{-}bcut}(B^*, h^*)}(h^*)$ we would like to query from $h^*$ in co-bcut$(B^*, h^*)$, but instead we perform a query from $\hat{h}^*$ in per$(B^*)$ to avoid constructing a data structure for co-bcut$(B^*, h^*)$.

We augment the arcs of $T(G)$ with the following during the construction of $B^*$: When we compute the extended shortest path trees from the hinges of $B^*$, we store $\mathrm{ecc}_{\mathrm{co\text{-}bcut}(B,h)}(h)$ with arc $h \rightarrow B$ in $T(G)$. These values are, at first, unknown for arcs in $T(G)$ on paths towards $B^*$. With each hinge vertex we store the largest and second largest known value among its adjacent arcs in $T(G)$ and two bags $B_1$ and $B_2$ attaining these values. How does this help us compute the maximum in (1)? When we learn of the missing value $\mathrm{ecc}_{\mathrm{co\text{-}bcut}(B^*,h^*)}(h^*)$ at $h^*$ in $T(G)$, we have three cases depending on whether $h^*$ has farthest points in the direction of $B^*$ or, if not, in the direction of $B'$. First, $\mathrm{ecc}_{\mathrm{co\text{-}bcut}(B^*,h^*)}(h^*)$ could be larger than $\mathrm{ecc}_{\mathrm{co\text{-}bcut}(B_1,h^*)}(h^*)$ in which case $\mathrm{ecc}_{\mathrm{bcut}(B',h^*)}(h^*) = \mathrm{ecc}_{\mathrm{co\text{-}bcut}(B^*,h^*)}(h^*)$. Second, $\mathrm{ecc}_{\mathrm{co\text{-}bcut}(B^*,h^*)}(h^*)$ could be strictly smaller than $\mathrm{ecc}_{\mathrm{co\text{-}bcut}(B_1,h^*)}(h^*)$ with $B_1 \neq B'$ in which case $\mathrm{ecc}_{\mathrm{bcut}(B',h^*)}(h^*) = \mathrm{ecc}_{\mathrm{co\text{-}bcut}(B_1,h^*)}(h^*)$. Third, $\mathrm{ecc}_{\mathrm{co\text{-}bcut}(B^*,h^*)}(h^*)$ could be strictly smaller than $\mathrm{ecc}_{\mathrm{co\text{-}bcut}(B_1,h^*)}(h^*)$ with $B_1 = B'$ in which case $\mathrm{ecc}_{\mathrm{bcut}(B',h^*)}(h^*) = \mathrm{ecc}_{\mathrm{co\text{-}bcut}(B_2,h^*)}(h^*)$. With the aforementioned bookkeeping, we can handle each of the three cases in constant time and consequently construct $\mathrm{per}(B')$ in time proportional to the size of $B'$.

With the above technique, we construct $\mathrm{per}(B')$ re-using the preprocessing from the construction of $\mathrm{per}(B^*)$ and the augmented tree structure. In the same way, we construct the perspectives from all other neighbors of $B^*$, then all perspectives from the neighbors of all neighbors of $B^*$ and so forth for an overall construction time of $O(n)$. We inherit the query times from the data structures of trees and uni-cyclic networks; in summary this yields the following theorem for eccentricity queries in cactus networks.

**Theorem 10.** *Let $G$ be a cactus network with $n$ vertices. There is a data structure with $O(n)$ size and construction time supporting eccentricity queries on $G$ in $O(1)$ time from branches and in $O(\log l)$ time from blocks of size $l$.*

### 3.2 Farthest-Point Queries

Consider a farthest-point query from a point $q$ in bag $B$, as illustrated in Figure 18. First, we perform a farthest-point query from $q$ in the farthest-point data structure of the perspective of $G$ from $B$. This query yields all farthest-points from $q$ inside $B$ and it returns the vertex $\hat{h}$ representing the bag-cut $\mathrm{bcut}(B,h)$ when $q$ has farthest points in $\mathrm{bcut}(B,h)$. Any path from $q$ to a farthest point $\bar{q}$ in $\mathrm{bcut}(B,h)$ passes through $h$, hence $\bar{q}$ is also farthest from $h$ in $\mathrm{bcut}(B,h)$. Recall that $\mathrm{bcut}(B,h)$ consists of the co-bag-cuts $\mathrm{co\text{-}bcut}(B',h)$ of the bags $B'$ neighboring $B$ at $h$. To find all farthest points from $q$ in $\mathrm{bcut}(B,h)$, we cascade the query into the co-bag-cuts $\mathrm{co\text{-}bcut}(B',h)$ of those neighbors $B'$ of $B$ that contain farthest points from $h$ in $\mathrm{bcut}(B,h)$, i.e., $\mathrm{ecc}_{\mathrm{co\text{-}bcut}(B',h)}(h) = \mathrm{ecc}_{\mathrm{bcut}(B,h)}(h)$. For the decision into which co-bag-cuts to cascade, we consult the eccentricity values stored with the arcs of the tree structure.

Now assume we cascade a query from $q$ into the co-bag-cut $\mathrm{co\text{-}bcut}(B',h)$, as in Figures 18c and 18d. To query for the farthest points from $h$ in co-bag-cut $\mathrm{co\text{-}bcut}(B',h)$, we perform a farthest-point query from $\hat{h}'$ in $\mathrm{per}(B')$, where $\hat{h}'$ is the vertex representing $\mathrm{bcut}(B',h)$ in $\mathrm{per}(B')$. The farthest points from $\hat{h}'$ in $\mathrm{per}(B')$ are also farthest from $h$ in $\mathrm{per}(B') - \hat{h}'h$, which corresponds to co-bcut $\mathrm{co\text{-}bcut}(B',h)$. Since $\hat{h}'$ is a pendant vertex in $\mathrm{per}(B')$, this query takes time proportional to the number of reported farthest points in $\mathrm{per}(B')$. The farthest points from $\hat{h}'$ in $\mathrm{per}(B')$ that lie in $B'$ are farthest points form the original query point $q$, and those farthest points from $\hat{h}'$ in $\mathrm{per}(B')$ that represent bag-cuts adjacent to $B'$ indicate where we have to continue our search for other farthest points from $q$. In this fashion, we propagate the query from $q$ to the bags of $G$ along paths to farthest points from $q$. However, we might visit $\Omega(n)$ bags before we reach one that contains a farthest point from $q$. We improve the query time by introducing shortcuts in the tree structure to bypass long chains of bags without farthest points.

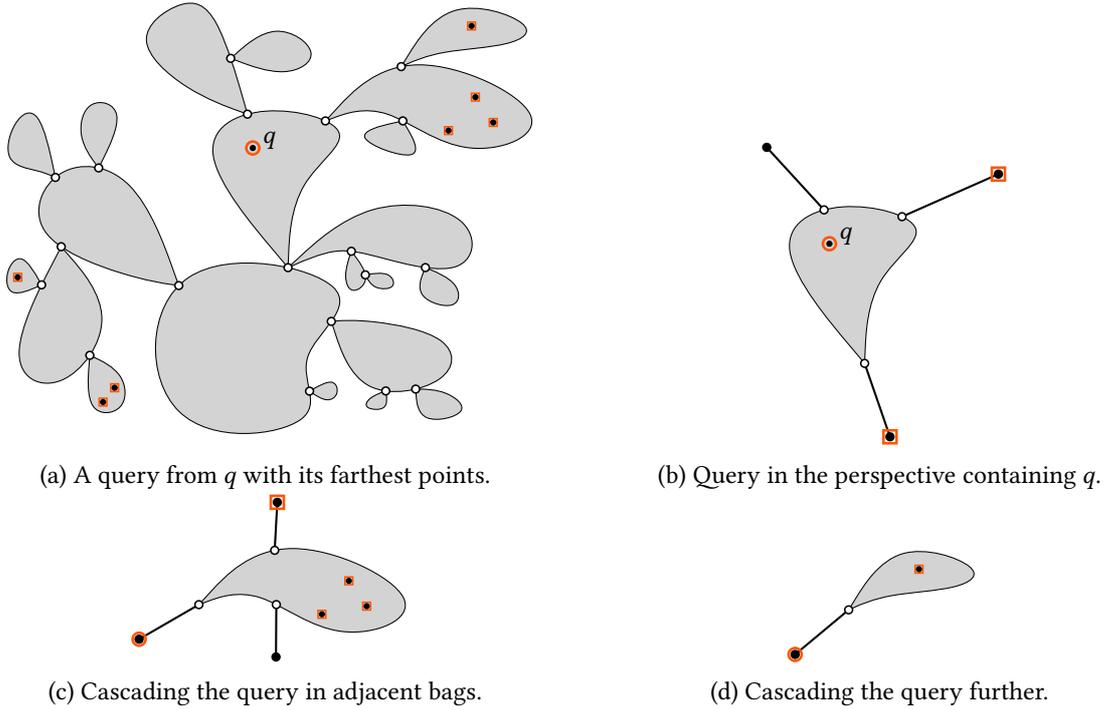

(a) A query from $q$ with its farthest points.

(b) Query in the perspective containing $q$.

(c) Cascading the query in adjacent bags.

(d) Cascading the query further.

Figure 18: Answering a farthest-point query from $q$ using the perspectives. Query points are indicated with orange circles, reported farthest points with orange squares. To process the query, we first query for $q$ in the perspective from the bag containing the query point (b). Then we cascade the query into adjacent bags that lead to further farthest points (c and d). Here, we show only two cascaded queries, to report all farthest points, we need five more cascaded queries.

Consider a co-bag-cut co-bcut$(B, h)$ of bag $B$ at hinge $h$. This co-bag-cut corresponds to the arc from $h$ to $B$ in the tree structure and all bags in co-bcut$(B, h)$ are in the sub-tree $T_{h \to B}$ reachable from $h$ through $B$. To give a visual idea of our shortcuts, imagine we do the following: First, we color a bag $B'$ in $T_{h \to B}$ red when $B'$ contains farthest points from $h$ in co-bcut$(B, h)$. Second, we color an uncolored bag $B'$ in $T_{h \to B}$ orange when two paths from $h$ to red bags split at $B'$. Finally, we color the remaining bags black. We seek to bypass black bags, since these are irrelevant for our query. For now, we only consider arcs of $T_{h \to B}$ leading away from $h$. The shortcut for arc $h' \to B'$ in $T_{h \to B}$ leads to the first arc $h'' \to B''$ where $B''$ is the closest orange or red descendant of $h'$. The shortcuts reachable from $h$ form a directed tree where all vertices representing bags are either yellow or red; following all shortcuts in this tree takes time $O(r)$ where $r$ is the number of red bag vertices in $T_{h \to B}$. In other words, using these shortcuts, the number of bags visited when reporting farthest points from $h$ in co-bcut$(B, h)$ is linear in the number of bags containing these farthest points. An example of a farthest-point query with and without using shortcuts is shown in Figure 19.

How can we determine these shortcuts efficiently? Recall that the propagation scheme starts at some bag $B^*$ where we first compute the extended shortest path tree of each hinge $h$ in $B^*$ in bcut$(B^*, h^*)$. We first consider only the arcs of $T(G)$ leading away from $B^*$. The arcs leading to leaf bags of $T(G)$ need no shortcuts. Consider an arc $h \to B$ and assume that all other arcs in the sub-tree $T_{h \to B}$ already have their shortcuts, as illustrated in Figure 20 on page 23. We distinguish three cases using a query in per$(B)$: First, there could be farthest points from $\hat{h}$ in $B$ (red case). Second, there could be farthest points from $\hat{h}$ in two bag cuts of $B$

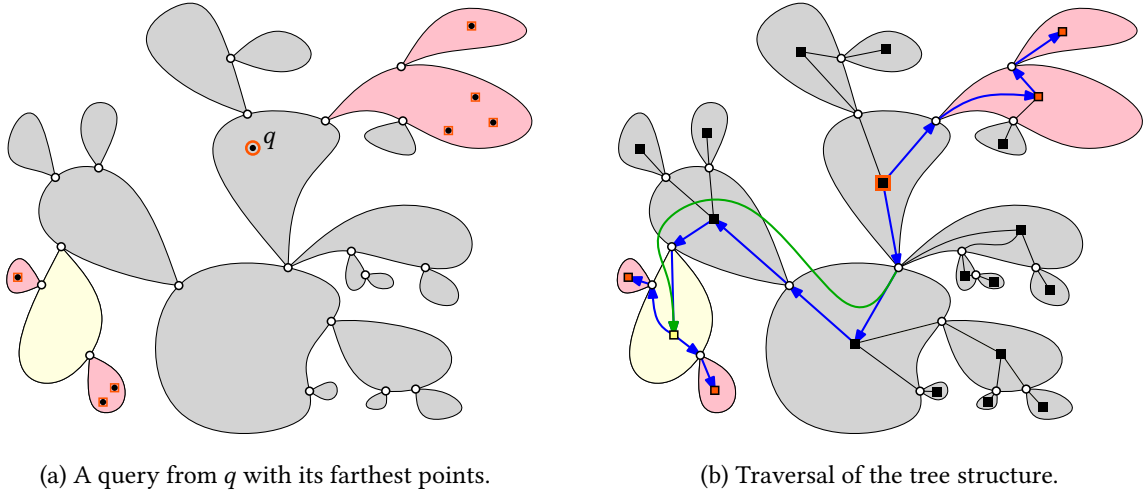

(a) A query from $q$ with its farthest points.  (b) Traversal of the tree structure.

Figure 19: The traversal of the tree strucuture (b) during a farthest-point query (a). The bags containing farthest-points are colored red, the bags where two paths to farthest points split are colored yellow, and all other blocks are colored gray. In blue, we highlight the arcs visited during the query and, in green, we highlight a shortcut bypassing several gray bags.

(orange case). Third, all farthest points from $\hat{h}$ could lie in a single bag cut bcut$(B, h')$ (black or orange case). In the first two cases, we introduce a trivial shortcut, i.e., the shortcut from $h \rightarrow B$ points to $h \rightarrow B$. In the third case, we inspect the information stored at the arcs incident to hinge $h'$ to determine whether the farthest points from $\hat{h}$ in bcut$(B, h')$ lie in multiple co-bag-cuts at $h'$ (orange case) or in a single co-bag-cut co-bcut$(B', h')$ (black case). In the black case, we introduce a shortcut from $h \rightarrow B$ to the destination of the shortcut from $h' \rightarrow B'$. In this way, we obtain all shortcuts in the tree structure leading away from $B^*$ without increasing our asymptotic bound on the size and construction time.

We employ our breadth-first-search propagation scheme to construct the shortcuts in the tree structure for arcs pointing towards $B^*$. At any hinge $h$ of $B^*$, we are only missing the shortcut for the arc $h \rightarrow B^*$. With a query in per$(B^*)$, we can immediately determine whether the red, the orange or the black case from above applies; no shortcuts towards $B^*$ are required for this step. With all shortcuts of the arcs from the hinges of $B^*$ in place, we can compute the shortcuts towards $B^*$ for the hinges of all blocks neighboring $B^*$, then all hinges of the bags neighboring the neighbors of $B^*$ and so forth.

Placing the shortcuts during the construction of our data structure for eccentricity queries takes only constant additional time and space per shortcut. How much time does a farthest point query take? First, we have to perform a farthest-point query in the perspective from the bag containing the query point. Then, we follow shortcuts towards bags containing more farthest points. The farthest-point queries in the perspective of subsequent bags $B'$ take time linear in the number of reported farthest points in per$(B')$, since we query from pendant vertices of per$(B')$. Moreover, as all visited blocks are either orange or red, the number of visited blocks is linear in the number of red blocks, i.e., blocks containing farthest points. Altogether, this yields a query time of $O(k)$ from branches and $O(k + \log l)$ from blocks of size $l$.

**Theorem 11.** *Let $G$ be a cactus network with $n$ vertices. There is a data structure with $O(n)$ size and construction time supporting farthest-point queries on $G$ in $O(k)$ time from branches and in $O(k + \log l)$ time from blocks of size $l$, where $k$ is the number of reported farthest points.*

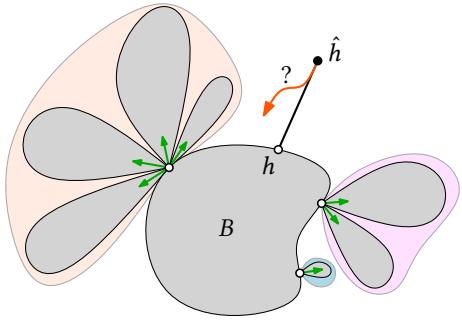

(a) The missing shortcut for co-bcut($B, h$).

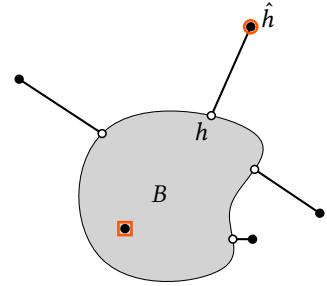

(b) We do not place a shortcut when the perspective from $B$ contains any farthest points within $B$ itself.

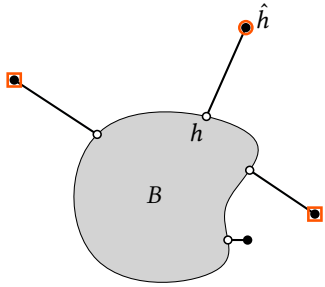

(c) We do not introduce a shortcut when there are farthest points in several adjacent bag-cuts.

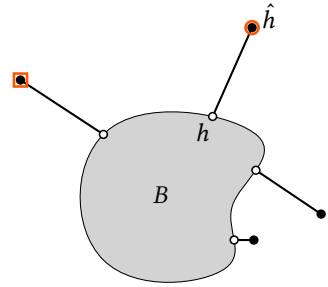

(d) When only one bag-cut contains farthest-points, we take a closer look at its co-bag-cuts.

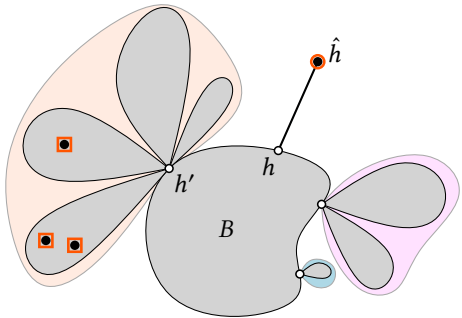

(e) We do not introduce a shortcut when several co-bag-cuts contain farthest points.

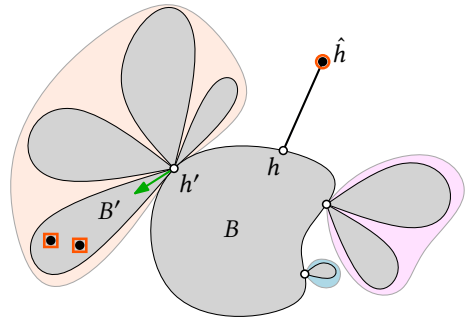

(f) We only introduce a shortcut when only one co-bag-cut contains farthest points.

Figure 20: Determining a missing shortcut into a co-bag-cut of $B$ at $h$, where the shortcuts for all co-bag cuts of the other hinges of $B$ are known (a). We consider the farthest points from $\hat{h}$ in per($B$): we place no shortcut when (b) there are farthest points in $B$, when (c) paths to farthest points split at $B$, and when (d,e) all farthest points lie in multiple co-bag-cuts of the same adjacent bag-cut. We only place a shortcut when (f) all farthest points lie in a single co-bag cut co-bcut($B', h'$). In this case, the shortcut from co-bcut($B, h$) leads to the target of the shortcut into co-bcut($B', h'$).

## 4 Conclusions and Future Work

In previous work [4], we obtain a data structure with construction time $O(m^2 \log n)$ for any network with $n$ vertices and $m$ edges, and with optimal query times for eccentricity queries and farthest-point queries. In this work, we improve the construction time to $O(n)$ for certain classes of networks without sacrificing optimal query time. In future work, we aim to achieve $o(m^2 \log n)$ construction time for more general classes of networks such as planar networks, $k$-almost-trees [10], and series parallel graphs [6].